\documentclass[ a4paper, 10pt, notitlepage, twocolumn, showpacs, aps, pra, amsmath, amsfonts, amssymb ]{revtex4-1}
\RequirePackage[utf8x]{inputenc}
\RequirePackage{braket}
\RequirePackage{graphicx}
\RequirePackage{color}
\definecolor{darkblue}{rgb}{0,0,.6}
\RequirePackage{hyperref} \hypersetup{colorlinks=true,linkcolor=darkblue,citecolor=darkblue,urlcolor=darkblue}
\begin{document}


\title{Entropic uncertainty relation for pointer-based simultaneous measurements of conjugate observables}
\author{Raoul Heese}
  \email{raoul.heese@uni-ulm.de}
  \affiliation{Institut f\"{u}r Quantenphysik, Universit\"{a}t Ulm - D-89069 Ulm, Germany}
\author{Matthias Freyberger}
  \affiliation{Institut f\"{u}r Quantenphysik, Universit\"{a}t Ulm - D-89069 Ulm, Germany}
\date{\today}

\begin{abstract}
We present a family of entropic uncertainty relations for pointer-based simultaneous measurements of conjugate observables. The lower bounds of these relations explicitly incorporate the influence of the measurement apparatus. We achieve this by using a mathematical theorem which states that the information entropy of convoluted probability distributions is bound from below. As a consequence of these results we can straightforwardly show that appropriately squeezed states are minimal entropy states for simultaneous measurements.
\end{abstract}

\maketitle


\section{Introduction}
The fundamental connection between any physical theory and phenomena in nature is being established by some kind of measurement. From this point of view, a profound understanding of measurement concepts is crucial to evaluating the validity of a physical theory. Especially in quantum mechanics, it is, however, not a simple task to gain a deeper knowledge of how measuring actually works and what it means ``to measure a system.'' There is a vast amount of literature on this topic from early works \cite{maxborn1926,heisenberg1927,schroedinger1935} up to more recent summaries \cite{braginsky1992,peres1998}. Our considerations are particularly founded on von~Neumann's pointer-based measurements \cite{vonneumann1968}, which treat the measurement apparatus as a quantum mechanical object called a \textit{pointer}. Von~Neumann states that from the interaction of the pointer and the system to be measured and a subsequent projective measurement of the pointer, one can deduce information about the system to be measured itself.\par
In this work we derive the entropic uncertainty \cite{bialynicki1975,deutsch1982,partovi1983,maassen1988} of a measurement configuration with two pointers bilinearly coupled to a system to be measured. Such a configuration allows a simultaneous measurement of two conjugate observables \cite{arthurs1965,stenholm1992}. In Sec.~\ref{sec:Concepts of pointer-based simultaneous measurements} we briefly review the underlying concept of this sort of pointer-based measurements. Subsequently, we discuss an entropic measure for the associated uncertainty in Sec.~\ref{Entropy as uncertainty measure of pointer-based simultaneous measurements}. In Sec.~\ref{Entropic uncertainty relations based on Lieb's theorem} we derive a family of entropic uncertainty relations, founded on a theorem of Lieb \cite{lieb1978}, and sort out the one which is best suited for our needs. Moreover, we show that this specific uncertainty relation is an improvement of a previously established entropic uncertainty relation \cite{wehrl1979, buzek1995}. Sec.~\ref{sec:Minimal entropy states} is dedicated to the derivation of minimal entropy states from our previous results. Finally, we conclude with a summary and an outlook in Sec.~\ref{Summary and Outlook}.


\section{Concepts of pointer-based simultaneous measurements}\label{sec:Concepts of pointer-based simultaneous measurements}
\begin{figure}[h]
  \includegraphics{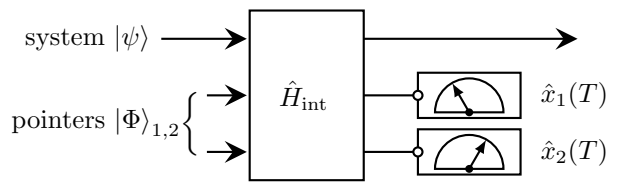}
  \caption{Schematic representation of a pointer-based measurement. The system to be measured, described by the state $\Ket{\psi}$, and the pointers, described by the bipartite state $\Ket{\Phi}_{1,2}$, interact with each other by means of an interaction Hamiltonian $\hat{H}_{\mathrm{int}}$. After the interaction process at time $T$, a projective measurement of the pointer positions $\hat{x}_{1}(T)$ and $\hat{x}_{2}(T)$ is performed. An appropriate choice of $\hat{H}_{\mathrm{int}}$ enforces that the measurable values of the pointer observables are related to the system's initial position and momentum. Hence, both of the latter can be retrieved from these measurement results. A fundamental question is therefore whether we can quantify the quantum uncertainty of such a simultaneous measurement by using the concept of an entropy.}
  \label{fig:measurement}
\end{figure}

As already loosely described in the Introduction, we consider a continuous-variable system which is coupled to two pointers, see Fig.~\ref{fig:measurement}. This coupling is chosen in such a way that the position of the system to be measured moves (or, in another phrasing, ``kicks'' \cite{busshardt2010}) the position of the first pointer whereas the corresponding momentum moves the position of the second pointer. We can therefore deduce information about the system from a projective measurement of the pointer positions after the coupling interaction. A more detailed discussion of this model can be found in Refs.~\cite{arthurs1965,stenholm1992}. We closely follow these references and just recall those results which are essential for the present study.\par
The total system is initially prepared in the state 
\begin{align} \label{eqn:State}
  \Ket{\Psi} \equiv \Ket{\psi} \otimes \Ket{\Phi}_{1,2},
\end{align}
where $\Ket{\psi}$ denotes the system to be measured and $\Ket{\Phi}_{1,2}$ represents the bipartite state of the two pointers. The coupling between system and pointers is described by the interaction Hamiltonian \cite{arthurs1965,stenholm1992} 
\begin{align} \label{eqn:Hamiltonian}
  \hat{H}_{\mathrm{int}} \equiv \kappa_1 \hat{x} \hat{p}_{1} + \kappa_2 \hat{p} \hat{p}_{2},
\end{align}
with constant coupling strengths $\kappa_1$ and $\kappa_2$ in an appropriate scaling (i.\,e.\ $\hbar = 1$) \cite{busshardt2010}. The operators $\hat{x}$ and $\hat{p}$ denote position and momentum of the system, respectively, whereas $\hat{p}_{1}$ and $\hat{p}_{2}$ indicate the momenta of either one of the two pointers.\par
It is the bilinear construction of the interaction Hamiltonian which is responsible for the behavior described above: The observable values of $\hat{x}$ and $\hat{p}$ move the positions of the first and second pointer, respectively. This is not surprising since Eq.~(\ref{eqn:Hamiltonian}) bears a strong resemblance to the displacement operator \cite{barnett1997}. However, the complete pointer dynamics are not immediately clear due to the mutual feedback of the three systems involved.

\subsection{Measurement coupling dynamics}
To understand the measurement coupling dynamics, we solve the corresponding Heisenberg equations for the positions $\hat{x}_1$ and $\hat{x}_2$ of the pointers. For this purpose we assume that the interaction energy is much larger than the kinetic energy (e.\,g., realized by a short but strong interaction between the system to be measured and the pointers) and we can therefore neglect free dynamics. For a given interaction time $T$, we arrive at the so-called inferred observables of position 
\begin{subequations} \label{eqn:InferredXP}
  \begin{align} \label{eqn:InferredX}
    \hat{\mathcal{X}} & \equiv \frac{\hat{x}_{1}(T)}{\kappa_1 T} = \frac{1}{\kappa_1 T} \exp[\operatorname{i} \hat{H}_{\mathrm{int}} T] \hat{x}_{1} \exp[-\operatorname{i} \hat{H}_{\mathrm{int}} T] \\
		      & = \hat{x} + \frac{\hat{x}_{1}}{\kappa_1 T} + \frac{\hat{p}_{2} \kappa_2 T}{2}
  \end{align}
and momentum
  \begin{align} \label{eqn:InferredP}
    \hat{\mathcal{P}} & \equiv \frac{\hat{x}_{2}(T)}{\kappa_2 T} = \frac{1}{\kappa_2 T} \exp[\operatorname{i} \hat{H}_{\mathrm{int}} T] \hat{x}_{2} \exp[-\operatorname{i} \hat{H}_{\mathrm{int}} T] \\
		      & = \hat{p} + \frac{\hat{x}_{2}}{\kappa_2 T} - \frac{\hat{p}_{1} \kappa_1 T}{2},
  \end{align}
\end{subequations}
which are just appropriately scaled pointer positions and clearly commute. Both $\hat{\mathcal{X}}$ and $\hat{\mathcal{P}}$ are Heisenberg operators which live in the combined Hilbert space of all three participating systems, Eq.~(\ref{eqn:State}). Observables without explicit time-dependencies refer to the initial time $t \equiv 0$ prior to the interaction process, i.\,e.\ observables in the Schr{\"o}dinger picture. Initial states of the pointers and therefore the expectation values $\braket{\hat{x}_{1}}$, $\braket{\hat{x}_{2}}$, $\braket{\hat{p}_{1}}$ and $\braket{\hat{p}_{2}}$ as well as the coupling strengths $\kappa_1$ and $\kappa_2$ are assumed to be known. Hence we conclude from Eq.~(\ref{eqn:InferredXP}) that one can deduce the expectation values of both the system's initial position $\braket{\hat{x}}$ and momentum $\braket{\hat{p}}$ by simultaneously determining the expectation values $\braket{\hat{\mathcal{X}}}$ and $\braket{\hat{\mathcal{P}}}$ of the pointer positions.\par
Then the interesting question arises as to how the uncertainties of these simultaneously measurable quantities behave. To find an answer, a description in terms of probability distributions for $\hat{\mathcal{X}}$ and $\hat{\mathcal{P}}$ is most appropriate. Therefore, we define the joint probability distribution
\begin{align} \label{eqn:ProbabilityXP}
  \mathrm{pr}(\mathcal{X},\mathcal{P}) \equiv \braket{ \Psi | \delta(\mathcal{X}-\hat{\mathcal{X}}) \delta(\mathcal{P}-\hat{\mathcal{P}}) | \Psi },
\end{align}
which describes the probability to infer a position $\mathcal{X}$, Eq.~(\ref{eqn:InferredX}), and a momentum $\mathcal{P}$, Eq.~(\ref{eqn:InferredP}), by means of a single pointer-based simultaneous measurement. In particular, Eq.~(\ref{eqn:ProbabilityXP}) directly results from the inverse Fourier transform of the corresponding characteristic function \cite{louisell1990} when we use the abbreviation
\begin{align}
  \delta(\mathcal{X}-\hat{\mathcal{X}}) \equiv \frac{1}{2 \pi} \int \limits_{- \infty}^{+ \infty} \! \mathrm d \alpha\ \exp[\operatorname{i} \alpha (\mathcal{X}-\hat{\mathcal{X}}) ]
\end{align}
and an analogous definition for $\delta(\mathcal{P}-\hat{\mathcal{P}})$. From another point of view, Eq.~(\ref{eqn:ProbabilityXP}) can be considered as the expectation value of a projection operator which corresponds to the joint eigenspace of the eigenvalues $\mathcal{X}$ and $\mathcal{P}$ of inferred position $\hat{\mathcal{X}}$ and inferred momentum $\hat{\mathcal{P}}$, respectively, and therefore projects the total state $\ket{\Psi}$, Eq.~(\ref{eqn:State}), onto the appropriate subspaces.\par
Integration of Eq.~(\ref{eqn:ProbabilityXP}) leads us to the marginal probability distribution of inferred position
\begin{subequations} \label{eqn:Probability}
  \begin{align} \label{eqn:ProbabilityX}
    \mathrm{pr}_{\mathcal{X}}(\mathcal{X}) \equiv \int \limits_{- \infty}^{+ \infty} \! \mathrm d \mathcal{P}\ \mathrm{pr}(\mathcal{X},\mathcal{P})
  \end{align}  
and inferred momentum
  \begin{align} \label{eqn:ProbabilityP}
    \mathrm{pr}_{\mathcal{P}}(\mathcal{P}) \equiv \int \limits_{- \infty}^{+ \infty} \! \mathrm d \mathcal{X}\ \mathrm{pr}(\mathcal{X},\mathcal{P}),
  \end{align}
\end{subequations}
respectively. Analogously, these distributions describe the probability to find either an inferred position $\mathcal{X}$ or an inferred momentum $\mathcal{P}$ through a single pointer-based measurement.

\subsection{The role of squeezed pointer states}\label{sec:The role of squeezed pointer states}
The concept of pointer-based measurements relies, as already mentioned in the beginning of Sec.~\ref{sec:Concepts of pointer-based simultaneous measurements}, on pointer states which are initially localized states in phase-space and can be kicked by the system state to imprint its information on them as a displacement in position. Therefore, it is reasonable to use vacuum states, or, more generally, squeezed states as initial pointer states. Historically, such states have already been used in Ref.~\cite{arthurs1965} in the pioneering theory of pointer-based measurements. Suitably squeezed pointer states also allow an optimal measurement apparatus in terms of variances \cite{busshardt2011} and we can connect our upcoming results to this statement in Sec.~\ref{sec:Minimal entropy states}.\par 
Hence, we assume in the following that our initial pointer states are squeezed vacuum states, which enables us to separate the initial bipartite state from Eq.~(\ref{eqn:State}) according to ${\ket{\Phi}_{1,2} \equiv \ket{\sigma_1}_{1} \otimes \ket{\sigma_2}_{2}}$. We confine the squeezing to real squeezing parameters without displacements \cite{barnett1997}. In other words, we only squeeze along the position and momentum axes in phase space. This approach allows us to define a squeezed state $\ket{\sigma}$ solely by its variance in position space ${\sigma^2 \equiv \braket{\sigma | \hat{x}^2 | \sigma }}$.\par
Due to these simplifications we can express the joint probability distribution, Eq.~(\ref{eqn:ProbabilityXP}), as a convolution
\begin{align} \label{eqn:ProbabilityXPSqueezed}
  \mathrm{pr}^{\mathrm{sq}}(\mathcal{X},\mathcal{P}) & = \frac{1}{2 \pi \delta_{\mathcal{X}} \delta_{\mathcal{P}}} \int \limits_{- \infty}^{+ \infty} \! \mathrm d x \! \int \limits_{- \infty}^{+ \infty} \! \mathrm d p\ W_{\ket{\psi}}(x, p) \nonumber \\
  & \ \ \ \ \ \times \exp \left[ - \frac{(\mathcal{X}-x)^2}{2 \delta_{\mathcal{X}}^2} - \frac{(\mathcal{P}-p)^2}{2 \delta_{\mathcal{P}}^2} \right]
\end{align}
of the Wigner function $W_{\ket{\psi}}$ of the state $\ket{\psi}$ and a Gaussian ``filter function'' \footnote{One can motivate a phase-space distribution which corresponds to Eq.~(\ref{eqn:ProbabilityXPSqueezed}), see S. Stenholm, \href{http://dx.doi.org/10.1088/0143-0807/1/4/012}{Eur. J. Phys. \textbf{1}, 244 (1980)}. A discussion of filter functions from an operational phase-space perspective can be found in K. W{\'o}dkiewicz, \href{http://dx.doi.org/10.1103/PhysRevLett.52.1064}{Phys. Rev. Lett. \textbf{52}, 1064 (1984)}. Furthermore, Eq.~(\ref{eqn:ProbabilityXPSqueezed}) is connected to the concept of propensities, see K. W{\'o}dkiewicz, \href{http://dx.doi.org/10.1016/0375-9601(86)90616-X}{Phys. Lett. A \textbf{115}, 304 (1986)}.}. The so-called noise terms \cite{busshardt2011}
\begin{subequations}\label{eqn:Noise}
  \begin{align} \label{eqn:NoiseX}
    \delta_{\mathcal{X}} \equiv \sqrt{\frac{\sigma_1^2}{\kappa_1^2 T^2} + \frac{\kappa_2^2 T^2}{16 \sigma_2^2}}
  \end{align}
and
  \begin{align} \label{eqn:NoiseP}
    \delta_{\mathcal{P}} \equiv \sqrt{\frac{\sigma_2^2}{\kappa_2^2 T^2} + \frac{\kappa_1^2 T^2}{16 \sigma_1^2}}
  \end{align}
\end{subequations}
describe the noise induced by the specific measurement setup. As written above, the expressions $\sigma_1^2$ and $\sigma_2^2$ stand for the initial variances in position of either one of the two pointer states (or, analogously, they stand for the respective second moments). One can easily verify that there is a minimal noise term product 
\begin{align} \label{eqn:MinimalNoiseTermProduct}
  \delta_{\mathcal{X}} \delta_{\mathcal{P}} \geq \frac{1}{2}
\end{align}
with equality for ${\sigma_1^2 \sigma_2^2 = \kappa_1^2 \kappa_2^2 T^4 / 16}$.\par
Using our joint probability distribution, Eq.~(\ref{eqn:ProbabilityXPSqueezed}), the marginal probability distributions, Eq.~(\ref{eqn:Probability}), take on the rather intuitive and simple forms \cite{busshardt2011}
\begin{subequations}\label{eqn:ProbabilitySqueezed}
  \begin{align} \label{eqn:ProbabilityXSqueezed}
    \mathrm{pr}_{\mathcal{X}}^{\mathrm{sq}}(\mathcal{X}) = \frac{1}{\sqrt{2 \pi}\delta_{\mathcal{X}}} \! \int \limits_{- \infty}^{+ \infty} \! \! \! \mathrm d x |\psi(x)|^2 \exp \left[ - \frac{(\mathcal{X}-x)^2}{2 \delta_{\mathcal{X}}^2} \right]
  \end{align}
and
  \begin{align} \label{eqn:ProbabilityPSqueezed}
    \mathrm{pr}_{\mathcal{P}}^{\mathrm{sq}}(\mathcal{P}) = \frac{1}{\sqrt{2 \pi}\delta_{\mathcal{P}}} \! \int \limits_{- \infty}^{+ \infty} \! \! \! \mathrm d p |\tilde{\psi}(p)|^2 \exp \left[ - \frac{(\mathcal{P}-p)^2}{2 \delta_{\mathcal{P}}^2} \right],
  \end{align}
\end{subequations}
respectively. Above we use the position representation ${\psi(x) \equiv \braket{x|\psi}}$ and the momentum representation ${\tilde{\psi}(p) \equiv \braket{p|\psi}}$ of the system state $\ket{\psi}$.\par
In the case of squeezed pointer states as discussed here, the influence of the setup on the simultaneous measurement is solely defined by the two noise terms of Eq.~(\ref{eqn:Noise}). We therefore do not need to consider specific coupling strengths, pointer state variances and interaction times in order to discuss different measurement setups in the following. Note, however, that different measurement setups can lead to the same noise terms and thus the same influence on the measurement.


\section{Entropy as uncertainty measure of pointer-based simultaneous measurements}\label{Entropy as uncertainty measure of pointer-based simultaneous measurements}
Variances are one possibility of describing the uncertainty of pointer-based measurements in an operational manner, but they are not always a good solution \cite{hilgevoord1990,bialynicki2011}. We take a step further away from these kinds of ``traditional uncertainty relations'' towards an entropic perspective.

\subsection{Collective Entropy}
The Shannon entropy has already been introduced \cite{buzek1995} as a measure of uncertainty in the context of pointer-based measurements. We start by closely following this concept and define the marginal entropy of the inferred position
\begin{subequations} \label{eqn:MarginalEntropy}
\begin{align} \label{eqn:MarginalEntropyX}
  S_{\mathcal{X}} \equiv - \int \limits_{- \infty}^{+ \infty} \! \mathrm d \mathcal{X}\ \mathrm{pr}_{\mathcal{X}}^{\mathrm{sq}}(\mathcal{X}) \ln \mathrm{pr}_{\mathcal{X}}^{\mathrm{sq}}(\mathcal{X})
\end{align}
and the marginal entropy of the inferred momentum
\begin{align} \label{eqn:MarginalEntropyP}
  S_{\mathcal{P}} \equiv - \int \limits_{- \infty}^{+ \infty} \! \mathrm d \mathcal{P}\ \mathrm{pr}_{\mathcal{P}}^{\mathrm{sq}}(\mathcal{P}) \ln \mathrm{pr}_{\mathcal{P}}^{\mathrm{sq}}(\mathcal{P})
\end{align}
\end{subequations}
based on the marginal probability distributions, Eq.~(\ref{eqn:ProbabilitySqueezed}). These entropies are connected to the uncertainty of a measurement of either inferred position or inferred momentum, respectively. Since an entropy describes the expected information gain, a large entropy and thus large expected information gain can be regarded as a high uncertainty, whereas a small entropy and thus small expected information gain can be regarded as a low uncertainty. Likewise,
\begin{align} \label{eqn:CollectiveEntropy}
  S \equiv S_{\mathcal{X}} + S_{\mathcal{P}}
\end{align}
is a suitable measure for the total uncertainty of a simultaneous measurement of inferred position and momentum and we therefore call it \emph{collective entropy}. For all these definitions we still assume squeezed pointer states.

\subsection{Entropic uncertainty relations}\label{sec:Entropic uncertainty relations}
From an operational point of view, one has no control over the system state to be measured. Instead, the system state is unknown and it is only possible to modify the measurement setup to obtain different sets of noise terms, Eq.~(\ref{eqn:Noise}). What can we say about the fundamental limits of an entropic uncertainty imposed by such a pointer-based simultaneous measurement setup?\par
To answer this question we need to introduce the concept of entropic uncertainty relations. Any entropic uncertainty relation of a pointer-based simultaneous measurement sets a lower bound for the collective entropy, Eq.~(\ref{eqn:CollectiveEntropy}), of the form \footnote{Differential entropies like the marginal entropies, Eq.~(\ref{eqn:MarginalEntropy}), do not have any lower bound in general. This is a consequence of the continuity of the associated probability distribution. As a result, every possible lower bound of the collective entropy must solely be based on the inner structure of the marginal probability distributions, Eq.~(\ref{eqn:ProbabilitySqueezed}).}
\begin{align} \label{eqn:EUR}
  S \geq \mathcal{S}_{\mathrm{min}}.
\end{align}
The \textit{quality} of such a lower bound $\mathcal{S}_{\mathrm{min}}$ for a given measurement setup (defined by the noise terms) and a particular system to be measured is determined by the difference of the left-hand side and the right-hand side of this inequality: the larger, the worse.\par
Uncertainty relations with a system-dependent right-hand side are not of much use if we maintain our operational assumption of an unknown system state \cite{deutsch1982,hilgevoord1990,bialynicki2011}. For this reason, we aim for an entropic uncertainty relation with a right-hand side which is independent of the system state to be measured. The measurement setup itself, since being known, is allowed to have an influence on our desired bound $\mathcal{S}_{\mathrm{min}}$.\par
Various versions of entropic uncertainty relations can already be found in the existing literature \cite{bialynicki1975,deutsch1982,partovi1983}. More recent studies mainly deal with sets of more than two observables in the context of quantum information theory \cite{wehner2010,bialynicki2011}. However, most results from these publications are not directly applicable to pointer-based simultaneous measurements. Therefore, we cannot further discuss all of them but rather concentrate on a pioneering approach for simultaneous measurements: In fact it was noted in Ref.~\cite{buzek1995} that the lower bound of the Wehrl entropy \cite{wehrl1979,lieb1978} can be associated with a constant lower bound of the collective entropy which reads
\begin{align} \label{eqn:Wehrl}
  S \geq \Omega \equiv 1 + \ln (2 \pi).
\end{align}
Thus, in this entropic uncertainty relation neither the system state to be measured nor the measurement setup have any influence on the lower bound $\Omega$ of the collective entropy. Due to this static behavior, it seems natural to search for a modification of Eq.~(\ref{eqn:Wehrl}) which incorporates the properties of the measurement configuration. And indeed, a whole family of such entropic uncertainty relations can be found based on a theorem by Lieb \cite{lieb1978}. The bounds of these uncertainty relations depend on two parameters and can effectively be increased beyond $\Omega$.\par
For example, we can obtain
\begin{align} \label{eqn:WehrlImproved}
  S \geq 1 + \ln \bigg[ 2 \pi \Big( \delta_{\mathcal{X}} \delta_{\mathcal{P}} + \frac{1}{2} \Big) \bigg] \geq \Omega.
\end{align}
Hence, for any noise term product ${\delta_{\mathcal{X}} \delta_{\mathcal{P}} > 1/2}$ this new bound is superior to $\Omega$. In the following section we discuss this improvement in its general form.


\section{Entropic uncertainty relations based on Lieb's theorem}\label{Entropic uncertainty relations based on Lieb's theorem}
In Ref.~\cite{lieb1978}, Lieb proves an inequality for the entropy of convolutions as a byproduct \footnote{In principle, Ref.~\cite{lieb1978} is centered around the proof of a lower bound of the Wehrl entropy \cite{wehrl1979}.} and remarks that it ``may be useful for related problems.'' Indeed, we can use his theorem in the following to establish a general lower bound of the collective entropy of pointer-based simultaneous measurements with squeezed pointer states.

\subsection{Lieb's theorem}\label{sec:Lieb's theorem}
The theorem \cite{lieb1978} states that if one has two non-negative and square-integrable functions $f$ and $g$ which are normalized according to
\begin{align}
  \int \limits_{- \infty}^{+ \infty} \! \mathrm d x\ f(x)= \int \limits_{- \infty}^{+ \infty} \! \mathrm d x\ g(x) = 1,
\end{align}
an information entropy of the form
\begin{align} \label{eqn:Entropy}
  S[ f ] \equiv - \int \limits_{- \infty}^{+ \infty} \! \mathrm d x\ f(x) \ln f(x)
\end{align}
obeys the inequality \footnote{There seems to be a minor notation inaccuracy in Lieb's original work \cite{lieb1978} as he defines entropies by including an additional factor $1/\pi$ on the right-hand side of Eq.~(\ref{eqn:Entropy}) but also arrives at Eq.~(\ref{eqn:LiebTheorem}). A proof of Eq.~(\ref{eqn:LiebTheorem}) can be accomplished by means of measure theory and will not be further discussed here.}
\begin{align} \label{eqn:LiebTheorem}
  S[ f * g ] \geq & \phantom{-..} \lambda S[ f ] + ( 1 - \lambda ) S[ g ] \nonumber \\
		  & - \frac{\lambda \ln \lambda + (1 - \lambda) \ln ( 1 - \lambda )}{2}.
\end{align}
Here the notation $f * g$ stands for the convolution
\begin{align}
  (f * g)(x) \equiv \int \limits_{- \infty}^{+ \infty} \! \mathrm d y\ f( y ) g( x - y )
\end{align}
and $\lambda$ is an arbitrary weighting parameter in the range ${0 \leq \lambda \leq 1}$.\par
Furthermore, we note that the equality in Eq.~(\ref{eqn:LiebTheorem}) holds true if and only if both $f$ and $g$ are Gaussians and one chooses \cite{lieb1978}
\begin{align} \label{eqn:GaussianParameter}
  \lambda_{\mathrm{G}}(\sigma_f,\sigma_g) \equiv \frac{\sigma_f^2}{\sigma_f^2+\sigma_g^2}
\end{align}
as the weighting parameter $\lambda$. Here, $\sigma_f^2$ and $\sigma_g^2$ denote the variances of $f$ and $g$, respectively.

\subsection{A family of entropic uncertainty relations}
Using the fact that the marginal probability distributions for squeezed pointer states, Eq.~(\ref{eqn:ProbabilitySqueezed}), are actually convolutions of the system state probability distribution with a Gaussian function, we can now directly apply Eq.~(\ref{eqn:LiebTheorem}) to both marginal entropies of pointer-based measurements, Eq.~(\ref{eqn:MarginalEntropy}). This results in a family of uncertainty relations for the collective entropy, Eq.~(\ref{eqn:CollectiveEntropy}), which reads
\begin{align}
  S \geq \Lambda(\lambda_{\mathcal{X}},\lambda_{\mathcal{P}})
\end{align}
with a parameterized lower bound
\begin{align} \label{eqn:Lieb}
  \Lambda(\lambda_{\mathcal{X}},\lambda_{\mathcal{P}}) \equiv & \phantom{+..} \lambda_{\mathcal{X}} S[|\psi|^2] + \lambda_{\mathcal{P}} S[|\tilde{\psi}|^2] \nonumber \\ 
  & + \frac{ 1 - \lambda_{\mathcal{X}} }{2} \ln ( 2 \pi \delta_{\mathcal{X}}^2 ) +  \frac{ 1 - \lambda_{\mathcal{P}} }{2} \ln ( 2 \pi \delta_{\mathcal{P}}^2 ) \nonumber \\ 
  & + \Theta(\lambda_{\mathcal{X}}) + \Theta(\lambda_{\mathcal{P}})
\end{align}
and the abbreviation
\begin{align} \label{eqn:Theta}
  \Theta(\lambda) \equiv \frac{ 1 - \lambda }{2} \left[ 1 - \ln ( 1 - \lambda ) \right] - \frac{\lambda}{2} \ln \lambda.
\end{align}
Here, a pair ${(\lambda_{\mathcal{X}},\lambda_{\mathcal{P}})}$ of weighting parameters occurs, which are both restricted to the interval $[0,1]$. Specifically, they describe the ratio of influence on the lower bound from either the system state to be measured or the measurement setup, represented by the noise terms, Eq.~(\ref{eqn:Noise}). Apart from these two contributions, correction terms $\Theta(\lambda)$, Eq.~(\ref{eqn:Theta}), occur in the lower bound, which are solely based on the weighting parameters and cannot directly be connected to a physical property. It is eventually the dependence on the measurement setup which makes Eq.~(\ref{eqn:Lieb}) superior in comparison with the constant bound, Eq.~(\ref{eqn:Wehrl}).

\subsection{Optimal weighting parameters}
In general, Eq.~(\ref{eqn:Lieb}) still depends on the system state to be measured. Yet, as already mentioned in Sec.~\ref{sec:Entropic uncertainty relations}, we strive for a lower bound which does not depend on the system state, so we need to find a pair of weighting parameters for which this dependence can be eliminated. Moreover, our choice should maximize Eq.~(\ref{eqn:Lieb}) as far as possible in order to improve our lower bound beyond the constant bound, Eq.~(\ref{eqn:Wehrl}).\par
Formally, we are looking for the optimal set of parameters $\lambda_{\mathcal{X}}^{\mathrm{O}}$ and $\lambda_{\mathcal{P}}^{\mathrm{O}}$ which lead to the largest and thus best lower entropy bound, Eq.~(\ref{eqn:Lieb}), by calculating an optimized bound
\begin{align} \label{eqn:LiebOptimized}
  \Lambda(\lambda_{\mathcal{X}}^{\mathrm{O}},\lambda_{\mathcal{P}}^{\mathrm{O}}) \equiv \underset{\lambda_{\mathcal{X}},\lambda_{\mathcal{P}} \in [0,1]}{\operatorname{max}} \Lambda(\lambda_{\mathcal{X}},\lambda_{\mathcal{P}}),
\end{align}
which should also be independent of the system state. To approach this problem, we first examine three critical pairs of weighting parameters: ${(1,1)}$, ${(0,0)}$ and ${(1/2,1/2)}$.\par
In case of the pair ${(1,1)}$, we have a purely system-determined bound
\begin{align} \label{eqn:LiebSystem}
  \Lambda(1,1) = S[|\psi|^2] + S[|\tilde{\psi}|^2].
\end{align}
Guided by our demand for a system-independent lower bound, we can further simplify Eq.~(\ref{eqn:LiebSystem}) by means of the entropic inequality \cite{hirschman1957,babenko1961,beckner1975,bialynicki1975,maassen1988}
\begin{align} \label{eqn:Hirsch}
  S[|\psi|^2] + S[|\tilde{\psi}|^2] \geq 1 + \ln \pi
\end{align}
to yield a constant bound. However, this resulting constant bound is worse than the constant bound from Eq.~(\ref{eqn:Wehrl}). This is understandable since Eq.~(\ref{eqn:Hirsch}) quantifies only an intrinsic uncertainty of the system. By no means does it takes into account that we perform a simultaneous measurement which will introduce additional uncertainties.\par
For the pair ${(0,0)}$ we arrive at a noise-determined bound \footnote{The noise-determined bound, Eq.~(\ref{eqn:LiebPointer}), can also be established by means of Jensen's inequality. From this point of view, it can be even further improved with the help of ``convexifications'', see S. Zlobec, \href{http://dx.doi.org/10.3336/gm.40.2.05 }{Glas. Mat. \textbf{40}, 241 (2005)}. However, we do not further pursue such considerations in this work.}
\begin{align} \label{eqn:LiebPointer}
  \Lambda(0,0) = 1 + \ln( 2 \pi \delta_{\mathcal{X}} \delta_{\mathcal{P}} ),
\end{align}
which is solely determined by measurement uncertainties. Although Eq.~(\ref{eqn:LiebPointer}) does not depend on the system state to be measured and is equal or better than the constant bound, Eq.~(\ref{eqn:Wehrl}), for a noise term product ${\delta_{\mathcal{X}} \delta_{\mathcal{P}} \geq 1}$, it is still worse for smaller noise term products.\par
Apparently, neither neglecting the influence of the measurement, Eq.~(\ref{eqn:LiebSystem}), nor neglecting the influence of the system, Eq.~(\ref{eqn:LiebPointer}), leads to a bound which is generally better than the constant bound, Eq.~(\ref{eqn:Wehrl}). Therefore, it may seem natural to choose the pair ${(1/2,1/2)}$ and we consequently arrive at a kind of ``balanced'' bound
\begin{subequations} \label{eqn:LiebBalanced}
\begin{align}
  \Lambda\left(\frac{1}{2},\frac{1}{2}\right) & = \frac{1 + S[|\psi|^2] + S[|\tilde{\psi}|^2] + \ln( 8 \pi \delta_{\mathcal{X}} \delta_{\mathcal{P}} )}{2} \\
		   & = \ln 2 + \frac{1}{2} \left[ \Lambda(0,0) + \Lambda(1,1) \right].
\end{align}
\end{subequations}
It contains the expressions from both Eqs.~(\ref{eqn:LiebSystem}) and (\ref{eqn:LiebPointer}). Moreover, when we further use Eq.~(\ref{eqn:Hirsch}) it is also possible to eliminate the intrinsic uncertainty of the system from Eq.~(\ref{eqn:LiebBalanced}). The resulting bound, whose uncertainty relation reads
\begin{align} \label{eqn:LiebBalancedSimplified}
  S \geq 1 + \ln ( 2 \pi \sqrt{2 \delta_{\mathcal{X}} \delta_{\mathcal{P}}} ),
\end{align}
is equal (${\delta_{\mathcal{X}} \delta_{\mathcal{P}} = 1/2}$) or better (${\delta_{\mathcal{X}} \delta_{\mathcal{P}} > 1/2}$) than the constant bound, Eq.~(\ref{eqn:Wehrl}).\par
Although this example already shows that Eq.~(\ref{eqn:Lieb}) is clearly an improvement over the constant bound, Eq.~(\ref{eqn:Wehrl}), we have not yet found the optimal pair of weighting parameters. Since our last example of equal weighting parameters has unfolded promising results, we use a general approach with arbitrary but equal weighting parameters in the following. Such a setting allows an analytical maximization of the lower bound and furthermore enables us to use Eq.~(\ref{eqn:Hirsch}) in order to eliminate the dependency on the system state.

\subsection{Single parameter bound}\label{Single parameter bound}
Let us consider the case of equal but arbitrary weighting parameters ${\lambda_{\mathcal{X}} \equiv \lambda_{\mathcal{P}} \equiv \lambda \in [0,1]}$. We can then directly simplify Eq.~(\ref{eqn:Lieb}) with the help of Eq.~(\ref{eqn:Hirsch}) and thus arrive at the single parameter bound
\begin{align} \label{eqn:LiebSingleParam}
  S \geq \Lambda_S (\lambda) \equiv 1 - \lambda \ln \frac{\lambda}{\pi} + (1-\lambda) \ln \left( \frac{2 \pi \delta_{\mathcal{X}} \delta_{\mathcal{P}} }{1-\lambda} \right).
\end{align}
This bound does not depend on the system state to be measured anymore. A maximization of ${\Lambda_S (\lambda)}$ leads us to the optimal single parameter bound associated with the entropic uncertainty relation
\begin{align} \label{eqn:LiebNoSystem}
  S \geq 1 + \ln \bigg[ 2 \pi \Big( \delta_{\mathcal{X}} \delta_{\mathcal{P}} + \frac{1}{2} \Big) \bigg],
\end{align}
which has already been presented in Eq.~(\ref{eqn:WehrlImproved}). As previously mentioned, Eq.~(\ref{eqn:LiebNoSystem}) is a refinement of the constant entropic bound, Eq.~(\ref{eqn:Wehrl}). Only in the case of a minimal noise term product ${\delta_{\mathcal{X}} \delta_{\mathcal{P}}=1/2}$ are both expressions identical. Moreover, Eq.~(\ref{eqn:LiebNoSystem}) is independent of the system state to be measured. This improved bound, Eq.~(\ref{eqn:LiebNoSystem}), is actually the main result of this contribution.


\section{Minimal entropy states}\label{sec:Minimal entropy states}
To deepen our understanding of the improved entropic uncertainty relation, Eq.~(\ref{eqn:LiebNoSystem}), we discuss it here in the context of so-called minimal entropy states. For a given measurement setup, defined by a set of noise terms, Eq.~(\ref{eqn:Noise}), minimal entropy states are those system states that result in a minimal collective entropy, Eq.~(\ref{eqn:CollectiveEntropy}). Although we have already emphasized that in operational terms one has no control over the system state to be measured, it is still interesting to know the ideal properties of a system state in a simultaneous pointer-based measurement. Moreover, any entropic uncertainty relation for a minimal entropy state is by definition also valid for any other state if the measurement setup remains unchanged. In other words, it is in fact the collective entropy of a minimal uncertainty state itself which sets the optimal bound.

\subsection{The single parameter bound from another perspective}
First, we take a look at the lower bound determining Eq.~(\ref{eqn:LiebNoSystem}) in the light of squeezed system states. In fact, we show that this bound is equal to the collective entropy of a particular squeezed system state, i.\,e.\ the inequality is tight. This squeezed system state is the one which results in minimal collective entropy for a given measurement setup.\par
We prove this statement in a straightforward way by utilizing Eq.~(\ref{eqn:GaussianParameter}) to determine a pair of weighting parameters ${(\lambda_{\mathrm{G}}(\sigma,\delta_{\mathcal{X}}),\lambda_{\mathrm{G}}(\sigma^{-1}/2,\delta_{\mathcal{P}}))}$ for a squeezed system state $\ket{\sigma}$ with variance $\sigma^2$ in position space (where the notation from section~\ref{sec:The role of squeezed pointer states} is being used). For this choice of parameters, the equality in Eq.~(\ref{eqn:LiebTheorem}) holds true and the collective entropy of a squeezed system state reads
\begin{align} \label{eqn:LiebGaussOpt}
  S[\ket{\sigma}] = 1 + \ln \pi - \ln \sqrt{ \lambda_{\mathrm{G}}(\sigma,\delta_{\mathcal{X}}) \lambda_{\mathrm{G}}(\sigma^{-1}/2,\delta_{\mathcal{P}})}.
\end{align}
To find the smallest collective entropy, Eq.~(\ref{eqn:CollectiveEntropy}), we can minimize Eq.~(\ref{eqn:LiebGaussOpt}) with respect to $\sigma$ and consequently arrive at
\begin{subequations} \label{eqn:LiebSqueezedEntropy}
\begin{align}
 \underset{\sigma}{\operatorname{min}}\ S[\ket{\sigma}] & = S[\ket{\sigma_{\mathrm{min}}}] \\
  & = 1 + \ln \bigg[ 2 \pi \Big( \delta_{\mathcal{X}} \delta_{\mathcal{P}} + \frac{1}{2} \Big) \bigg]
\end{align}
\end{subequations}
with \footnote{Eq.~(\ref{eqn:MinimalVariance}) is not only a condition for minimal entropy but also a condition for minimal variance \cite{busshardt2011}. This connection is not surprising since we have used squeezed states as pointer states and thus a Gaussian filter function, Eq.~(\ref{eqn:ProbabilityXPSqueezed}).}
\begin{align} \label{eqn:MinimalVariance}
  \sigma_{\mathrm{min}}^2 \equiv \frac{\delta_{\mathcal{X}}}{2 \delta_{\mathcal{P}}}.
\end{align}
Hence we see that the collective entropy of the corresponding squeezed states reaches the lower bound of our entropic uncertainty relation, Eq.~(\ref{eqn:LiebNoSystem}). From this relationship one can easily argue that squeezed states are minimal entropy states if they obey Eq.~(\ref{eqn:MinimalVariance}).

\subsection{A family of minimal entropy states}
The collective entropy of any system state $\ket{\psi}$ is limited from below by the uncertainty relation, Eq.~(\ref{eqn:LiebNoSystem}). On the other hand, according to Eq.~(\ref{eqn:LiebSqueezedEntropy}), the corresponding lower bound is also equivalent to the collective entropy of a squeezed system state $\ket{\sigma_{\mathrm{min}}}$ with the specific variance $\sigma_{\mathrm{min}}^2$, Eq.~(\ref{eqn:MinimalVariance}), in position space. Therefore, one has ${S[\ket{\psi}] \geq S[\ket{\sigma_{\mathrm{min}}}]}$ and thus, we can deduce that minimal collective entropy can always be reached by a squeezed system state $\ket{\sigma_{\mathrm{min}}}$ whose variance in position space fulfills Eq.~(\ref{eqn:MinimalVariance}). Consequently, the bound given by our entropic uncertainty relation, Eq.~(\ref{eqn:LiebNoSystem}), is the optimal bound for a simultaneous measurement based on squeezed pointer states.\par
In fact, since the equality in Eq.~(\ref{eqn:LiebTheorem}), is fulfilled if and only if one has Gaussian probability distributions which obey Eq.~(\ref{eqn:GaussianParameter}), the minimal entropy states $\ket{\sigma_{\mathrm{min}}}$ are (apart from a global phase factor) unique minimal entropy states for a given measurement setup. This means that there may not be a differently shaped state ${\ket{\psi} \neq \alpha \ket{\sigma_{\mathrm{min}}}}$ ({$|\alpha| = 1$}) which also reaches minimal collective entropy \footnote{This statement can be verified numerically by composing $\ket{\psi}$ as a superposition of number states and optimizing the resulting collective entropy, Eq.~(\ref{eqn:CollectiveEntropy}), with respect to the coefficients of the superposition by means of a Nelder-Mead simplex algorithm (see, e.\,g., J. C. Lagarias, J. A. Reeds, M. H. Wright, and P. E. Wright, \href{http://dx.doi.org/10.1137/S1052623496303470}{SIAM J. Optim. \textbf{9}, 112 (1998)}). The resulting states are also squeezed states which expectedly fulfill Eq.~(\ref{eqn:MinimalVariance}).}.\par
We remark that the lowest possible collective entropy of a pointer-based simultaneous measurement with squeezed pointer states, i.\,e.\ ${S = 1+\ln(2\pi)}$, can be reached with a squeezed system state which obeys Eq.~(\ref{eqn:MinimalVariance}) and a measurement configuration with a minimal noise term product ${\delta_{\mathcal{X}} \delta_{\mathcal{P}}=1/2}$. The constant bound, Eq.~(\ref{eqn:Wehrl}), is apparently equivalent to this lowest limit. This result corresponds to the derivation of Eq.~(\ref{eqn:Wehrl}), where it is assumed that ``the filter state is considered to be in a pure minimum uncertainty state'' \cite{buzek1995}.


\section{Summary and Outlook}\label{Summary and Outlook}
In this final section we outline the most important results of this contribution and take a look at several aspects which still leave room for further considerations.

\subsection{Summary}
We have analyzed a whole family of entropic uncertainty relation for pointer-based simultaneous measurements. This family includes specific cases which have been known previously. However, it has also allowed us to eliminate the influence of the system state to be measured from the bound while not neglecting the influence of the measurement setup. This has led us to the optimal bound which is equal to the collective entropy of minimal entropy states. Such minimal entropy states are the well-known squeezed vacuum states.

\subsection{Outlook}
In the previous sections we imposed several limitations in order to simplify our mathematical expressions. In particular, we chose a linear interaction Hamiltonian with constant coupling strengths, squeezed pointer states and pure system states to describe the simultaneous measurement. Furthermore, we neglected free dynamics of all systems involved. In the following we discuss which of these limitations may be loosened.\par
Linearity of the interaction Hamiltonian is a crucial point of our discussion. On the other hand, we expect that higher order interaction terms only lead to corrections which are smaller than the main contribution of the linear term. Therefore, our coupling model should be able to describe the key features of pointer-based measurements and consequently allows us to discuss the associated noise of the measurement apparatus in terms of entropies. It is, however, possible to generalize our considerations to time-dependent coupling strengths, which results in a more complicated expression for the noise terms \cite{busshardt2010}. Otherwise, the discussion of the results remains unchanged. Similarly, all the effects of free motion in the Hamiltonian can be included in modified noise terms \footnote{The influence of free dynamics can be calculated with the help of characteristic functions if we use squeezed pointer states. An additional correlation of inferred position and inferred momentum appears in the joint probability distribution, Eq.~(\ref{eqn:ProbabilityXP}), in comparison with Eq.~(\ref{eqn:ProbabilityXPSqueezed}), but the resulting marginal probability distributions can nevertheless be written as in Eq.~(\ref{eqn:ProbabilitySqueezed}) with modified noise terms $\delta_{\mathcal{X}}$ and $\delta_{\mathcal{P}}$, cf. Eq.~(\ref{eqn:Noise}).}.\par 
Using general pointer states instead of squeezed pointer states prevents us from applying Lieb's theorem and we have to fall back onto different entropic uncertainty relations. For example, we could use a linearization approach to find an entropic bound, but there are also other possibilities \footnote{A promising approach is based on the continuous expansion of ${- \Sigma_k p_k \ln p_k \geq -\ln \operatorname{max} p_k}$ (with $\Sigma_k p_k = 1$), and is, e.\,g., valid due to the monotonicity of R{\'{e}}nyi entropies, see A. R{\'{e}}nyi, in \textit{Proceedings of the Fourth Berkeley Symposium on Mathematical Statistics and Probability}, Vol. 1 (University of California, Berkeley, 1961) pp. 547--561. It states that ${S \geq - \ln \left[ \operatorname{max}\mathrm{pr}_{\mathcal{X}}^{\mathrm{sq}}(\mathcal{X}) \cdot \operatorname{max}\mathrm{pr}_{\mathcal{P}}^{\mathrm{sq}}(\mathcal{P}) \right]}$. A recent attempt to express uncertainty relations in terms of the R{\'{e}}nyi entropy can be found in I. Bia{\l}ynicki-Birula, \href{http://dx.doi.org/10.1103/PhysRevA.74.052101}{Phys. Rev. A \textbf{74}, 052101 (2006)} and Ref.~\cite{bialynicki2011}.} which are valid for arbitrary pointer states.\par 
Lastly, extending our framework to mixed states for system and pointers requires us to rewrite most definitions and therefore cannot be further discussed here. The underlying limitation is related to the fact that we use a projective measurement to read out the pointer observables. A more physical approach would incorporate an environment to which the pointers are coupled \cite{zurek2003}. By doing so, the pointer observables could emerge purely by decoherence effects. Density matrices instead of pure states would appear naturally. Consequently, additional environmental noises are assumed to influence the measurement.


%

\end{document}